\documentclass[manuscript]{aastex}

\usepackage{color}
\usepackage{multicol}
\usepackage{graphicx}

\shorttitle{H-CO Collisional Quenching}
\shortauthors{Walker et al.}

\begin{document}

\title{Quantum Calculation of Inelastic CO Collisions with H. II.
Pure Rotational Quenching of High Rotational Levels}

\author{Kyle M. Walker\altaffilmark{1}, L. Song\altaffilmark{2},
 B. H. Yang\altaffilmark{1}, G. C. Groenenboom\altaffilmark{2},
 A. van der Avoird\altaffilmark{2}, N. Balakrishnan\altaffilmark{3},
 R. C. Forrey\altaffilmark{4}, 
 and P. C. Stancil\altaffilmark{1}}

\altaffiltext{1}{Department of Physics and Astronomy and Center for
 Simulational Physics, The University of Georgia, Athens, GA 30602, USA}
\altaffiltext{2}{Theoretical Chemistry, Institute for Molecules and
 Materials, Radboud University, Heyendaalseweg 135, 6525 AJ
 Nijmegen, The Netherlands}
\altaffiltext{3}{Department of Chemistry, University of Nevada
 Las Vegas, NV 89154, USA}
\altaffiltext{4}{Department of Physics, Penn State University, Berks
 Campus, Reading, PA 19610, USA}
\begin{abstract}

Carbon monoxide is a simple molecule present in many astrophysical environments, 
 and collisional excitation rate coefficients due to the dominant collision 
 partners are necessary to accurately predict spectral line intensities
 and extract astrophysical parameters.
 We report new quantum scattering calculations for rotational deexcitation 
 transitions of CO induced by H using the three-dimensional potential energy 
 surface~(PES) of \citet{song}.
 State-to-state cross sections for collision energies from 10$^{-5}$ to 15,000~cm$^{-1}$
 and rate coefficients for temperatures ranging from 1 to 3000~K are obtained for 
 CO($v=0$, $j$) deexcitation from $j=1-45$ to all lower $j'$ levels, where $j$ is the
 rotational quantum number. 
 Close-coupling and coupled-states calculations were performed in full-dimension for 
 $j=1-5$, 10, 15, 20, 25, 30, 35, 40, and 45
 while scaling approaches were used to estimate rate coefficients for all 
 other intermediate rotational states.
 The current rate coefficients are compared with previous scattering results using earlier
 PESs. Astrophysical applications of the current results are briefly discussed.  

\end{abstract}

\keywords{ISM: molecules --- molecular data --- molecular processes ---  photon-dominated region}
 
\section{Introduction}

Carbon monoxide is the second most abundant molecule in the universe
 after molecular hydrogen and is found in a variety of astrophysical environments.
 The formation of CO in low-density interstellar clouds proceeds mainly through
 gas-phase chemical reactions involving H$_2$, and therefore detecting and
 measuring CO spectral lines is one way to trace the molecular interstellar
 medium~(ISM) in the region. While H$_2$ rotational lines are difficult
 to observe from the ground,
 CO is more easily detected as its rovibrational transitions can be observed as 
 absorption in the ultraviolet~(UV) and near-infrared~(NIR) and emission in the NIR,
 far-infrared~(FIR), and submillimeter. The rotational energy level spacings 
 of a diatomic molecule depend inversely on its moment of inertia, and the relatively large
 moment of inertia of CO compared to H$_{2}$ yields rotational levels
 with small energy separation. The lowest rotational transition of CO has a wavelength of 
 $\sim$2.6~mm which gives an excitation temperature of just $\sim$5.5~K. 
 Therefore, CO can be collisionally excited to high rotational levels in photon-dominated regions
 (PDRs), in shocks, and in other moderately energetic environments.

A number of CO pure rotational lines starting at $j=14\rightarrow 13$, where $j$ is the 
 rotational quantum number, have been detected with 
 the Long-Wavelength Spectrometer~(LWS) \citep{clegg96} aboard the {\it Infrared
 Space Observatory} \citep[\textit{ISO},][]{kess96}. 
 These include the FIR spectra of all the lines up to $j=39\rightarrow 38$ in the  
 carbon-rich circumstellar envelope IRC +10216 \citep{cern96},  
 transitions up to $j=25\rightarrow 24$ in VY Canis Majoris and other oxygen-rich 
 circumstellar envelopes \citep{pole10},
 transitions up to $j=24\rightarrow 23$ from the carbon-rich planetary nebula 
 NGC~7027 \citep{liu96}, and transitions 
 up to $j=19\rightarrow 18$ of the Herbig Haro objects HH 52-53-54 and 
 IRAS 12496-7650 in a nearby star-forming region \citep{nis96}.
 The KOSMA 3m and IRAM 30m telescopes in Switzerland and Spain,
 respectively, performed observations of the low-$j$ transitions of 
 CO in the Rosette Molecular Complex \citep{sch98}, while 
 the {\it Odin Orbital Observatory} \citep{nord03} detected $j=5\rightarrow 4$ emission 
 from the PDR of Orion KL \citep{per07}.
 More recently, the \textit{Herschel Space Observatory} \citep{pilb10} 
 observed the $j=9\rightarrow 8$ line towards Monoceros R2 \citep{pill12} 
 and high-$j$ CO lines in the NGC 1333 low-mass star-forming 
 region \citep{yil10} with the
 Heterodyne Instrument for the Far-Infrared \citep[HIFI,][]{deg10}.
 The Spectral and Photometric Imaging REceiver \citep[SPIRE,][]{grif10}
 aboard {\it Herschel} has also probed the submillimeter molecular 
 interstellar medium of M82 from $j=4\rightarrow 3$ up to $j=13\rightarrow 12$ \citep{pan10}.

The chemical and physical conditions where CO resides are deduced from 
 observed spectral lines, but it is often not appropriate to assume local 
 thermodynamic equilibrium~(LTE) when modeling these regions. 
 In low-density environments where the level 
 populations routinely depart from LTE, collisional excitation 
 rate coefficients with the dominant species --- 
 mainly H$_2$, H, He, and electrons --- are necessary to accurately
 predict spectral line intensities.
 Although the effect of electron collisions is minor, 
 the role of collisional excitation of CO by H, especially 
 in environments such as diffuse molecular clouds \citep{gold13}
 and cool mixed atomic and molecular hydrogen gas \citep{lis06},
 cannot be neglected.
 Rate coefficient calculations for pure rotational excitation of the 
 first 15 rotational levels of the He-CO system were performed by 
 \citet{cecchi02} for the temperature range 5 to 500~K. 
 \citet{yang10} calculated reliable H$_{2}$-CO rate coefficients 
 for rotational transitions in CO induced by both para- and 
 ortho-H$_{2}$ collisions for deexcitation from $j=1-40$ to all 
 lower $j'$ levels for temperatures between 1 and 3000~K.
 Although the H-CO system has been studied extensively with the most 
 recent rate coefficients reported by \citet{yang13}, there is
 unsatisfactory agreement between the results obtained upon the various 
 potential energy surfaces~(PESs).

The earliest analysis of H-CO collisions was performed by 
 \citet{chu75}, who used a
 short-range, semi-empirical potential joined with a long-range 
 Buckingham potential to calculate cross sections via 
 two close-coupling formulations.
 The resulting cross sections were found to be comparable to 
 those of H$_2$-CO for small $\Delta j$. The Maxwellian-averaged 
 rate coefficients were calculated for the temperature range of 
 $5-150$~K.
 The following year, quantal calculations were carried out by 
 \citet{green76} over the same temperature range, but with a 
 new semi-empirical potential and strikingly different results.
 These collisional rate coefficients 
 were typically an order of magnitude smaller than those of 
 \citet{chu75}, and small $\Delta j$ transitions, 
 i.e., $|\Delta j|$ $=$ 1 or 2,
 were reported to have much larger rate coefficients when 
 compared to large $\Delta j$ transitions.
 The small magnitude of the H-CO rate coefficients obtained by 
 \citet{green76} led to the neglect of H as a collider in CO 
 emission modeling. 

The next interaction PES for the H-CO system followed a more 
 sophisticated level of theory using \textit{ab initio} calculations 
 for the surface \citep[][hereafter called BBH]{bow86}.
 These calculations employed the \citet{dun71} valence double-zeta 
 contractions of the \citet{huz71}, (9s,5p) sets
 of carbon- and oxygen-centered primitive Gaussians and used 
 the (4s,2s) contraction with a scale factor of 1.2 for hydrogen.
 Restricted Hartree-Fock calculations were carried out, followed by 
 configuration interaction computations including all 
 singly- and doubly-excited configurations. 
 The resulting empirically corrected potential 
 was used in dynamics calculations using a cubic spline interpolation.
 Subsequent coupled-channel scattering calculations were carried 
 out on this PES by \citet{lee87} who first noticed the propensity 
 for even-$\Delta j$ transitions for inelastic scattering.

Another surface was constructed by \citet{wern95}, hereafter WKS,
 to see if new experimental results could be more closely reproduced.
 The \textit{ab initio} electronic structure calculations were performed 
 with the MOLPRO program package \citep{mol} using the 
 internally contracted multireference 
 configuration interaction (icMRCI) method. 
 Resonance energies and widths from the WKS surface 
 agreed with experimental data \citep{kell96},
 but striking differences were apparent in the collision cross sections 
 obtained using the BBH and WKS surfaces \citep{green96}.
 The cross sections on the newer WKS surface, especially for 
 low-$j$ pure rotational transitions, yielded higher values 
 than those calculated on the BBH surface.
 \citet{bala02} obtained similar results when they used both 
 the BBH and the WKS surfaces to compute collisional rate coefficients 
 for temperatures in the range 5 to 3000~K.
 The discrepancy in the CO($j=1\rightarrow 0$) transition was the largest.
 However, \citet{bala02} performed explicit quantum-mechanical 
 scattering calculations with the WKS PES for the first eight
 pure rotational transitions (using the close-coupling [CC] framework)
 and first five vibrational levels (in the infinite order sudden~[IOS]
 approximation),
 and this complete set of rate coefficients has been extensively 
 used in astrophysical models.

\citet{shep07} revisited H-CO scattering by introducing two
 new rigid-rotor PESs and computing cross sections on each surface 
 at collision energies of 400 and 800~cm$^{-1}$.
 One \textit{ab initio} surface was calculated using the coupled cluster 
 method with single and double excitations and a perturbative 
 treatment of triple excitations [CCSD(T)] employing the frozen 
 core approximation \citep{purv82}. 
 Restricted Hartree-Fock orbitals were used for the open-shell
 calculations and the spin-restrictions were relaxed
 in the solution of the coupled cluster equations [R/UCCSD(T)]
 \citep{know93}. 
 The other PES was computed using the complete active space 
 self-consistent field~(CASSCF) \citep{knwe85} and icMRCI 
 \citep{wekn88} method with the aug-cc-pVQZ basis set for 
 H, C, and O by \citet{woon94} as implemented in the MOLPRO program. 
 Their results for both potentials are more similar to 
 \citet{green76} and it was recommended that the rate 
 coefficients calculated by \citet{bala02} using the inaccurate 
 WKS surface be abandoned.
 Based on this recommendation, astrophysical models using 
 the overestimated rate coefficients should also be reexamined.
 The most recent set of H-CO calculations were performed by \citet{yang13}
 and extend the calculations of \citet{shep07} on the MRCI
 rigid-rotor PES.
 State-to-state rotational deexcitation cross sections and rate 
 coefficients from initial states $j=1-5$ in the ground vibrational 
 state to all lower $j'$ levels were computed.
 
In this work, we report new quantum scattering calculations for 
 rotational deexcitation transitions of CO induced by H using the 
 three-dimensional~(3D) interaction PES of \citet{song}.
 This 3D \textit{ab initio} PES uses the spin-unrestricted open-shell single and 
 double excitation coupled-cluster method with perturbative triples [UCCSD(T)] 
 with molecular orbitals from restricted Hartree-Fock calculations~(RHF).
 Electronic structure calculations were performed with the MOLPRO 2000 package
 \citep{mol} and 3744 interaction energies were included in the fit.
 When compared to previous PESs, the new 3D surface uses the most 
 sophisticated level of theory and is the most accurate available 
 for scattering calculations.
 State-to-state cross sections for collision energies from 
 10$^{-5}$ to 15,000~cm$^{-1}$ are computed for CO($v=0$, $j$) deexcitation 
 from initial states $j=1-45$ to all lower $j'$ levels.
 While close-coupling and coupled-states calculations, in full-dimension, are performed 
 for $j=1-5$, 10, 15, 20, 25, 30, 35, 40, and 45, scaling approaches are used to estimate the 
 rate coefficients for all other intermediate rotational states.
 Rate coefficients for temperatures ranging from 1 to 3000~K are evaluated 
 and compared with previous scattering results obtained on earlier surfaces.
 The astrophysical implications of new rotational deexcitation rate
 coefficients are illustrated.

\section{Computational Approach}

We performed inelastic scattering calculations on the most current three-dimensional 
 PES of the H-CO complex in the ground state ($\widetilde{X}~^{2}A'$) \citep{song}.
 Computations were carried out using the quantum mechanical close-coupling (CC)
 method \citep{art60} for kinetic energies below 1000~cm$^{-1}$.
 From 1000~cm$^{-1}$ to 15,000~cm$^{-1}$ the coupled-states (CS) approximation 
 of \citet{mcg74} was utilized.
 We treated hydrogen as a structureless atom and allowed the bond length to vary for CO.
 The full 3D surface is used in the calculations; at no point is
 the CO bond length fixed nor a rigid-rotor approximation made.
 The interaction potential was expressed as $V$($R$,$r$,$\theta$),
 where $r$ is the CO intramolecular distance,
 $R$ the distance from the CO center of mass to the H nucleus,
 and $\theta$ the angle between the vector $R$ and the CO bond axis,
 where linear C-O-H has an angle of zero degrees and $\theta=180^\circ$ for linear H-C-O.
 The potential was expanded according to 
 \begin{equation} \label{potenl1}
 V(R,r,\theta ) = 
 \sum\limits_{\lambda =0}^{\lambda_{\rm max}} \upsilon_{\lambda }(R,r) 
 P_{\lambda }(\cos \theta ){\rm ,}
 \end{equation}
 where $P_{\lambda}$ are Legendre polynomials of order $\lambda$.
 The radial dependence of the potential used a 20-point Gauss-Hermite
 quadrature over $r$ to represent CO stretching.
 The angular dependence of the potential was expanded to 
 $\lambda_{\rm max}=20$ with a 22-point Gauss-Legendre quadrature.
 
The quantum-mechanical close-coupling calculations were performed
 using the mixed-mode OpenMP/MPI version of the nonreactive scattering program 
 MOLSCAT \citep{hut94} modified by \citet{val08} and \citet{walk13}.
 The modified log-derivative Airy propagator of \citet{alex87}
 with variable step size was used to solve the coupled-channel 
 equations.
 The propagation was carried out from $R=1$~a$_{0}$ to a 
 maximum distance of $R=100$~a$_{0}$, where a$_{0}$ is the atomic 
 unit of length (the Bohr radius).
 For calculations with initial state $j<25$, the basis set 
 included 31 rotational levels in the ground vibrational state of CO.
 For initial states $j\ge 25$ the basis sets 
 included at least $5-10$ closed rotational levels.
 Only pure rotational transitions within the $v=0$ vibrational level
 are reported here.
 A convergence study was performed with excited vibrational levels in
 the basis set, but the addition of higher vibrational levels yielded
 results within 2\%.
 A sufficient number of angular momentum partial waves were included 
 to ensure convergence to within 10\% for the largest state-to-state cross sections.
 For the low energy range, 10$^{-5}$--10~cm$^{-1}$, the number of maximum partial waves, $J_{\rm max}$,
 was increased each decade resulting in the values 2, 4, 8, 12, 14, and 20 being 
 added to the value of initial $j$, 
 respectively, while
 converged cross sections for energies from 
 $10-49$, $50-95$, $100-900$, $1000-9000$, and $10,000-15,000$~cm$^{-1}$ 
 were calculated by adding the values of 26, 30, 40, 50, and 60 to
 that of the initial state, respectively.

The degeneracy-averaged-and-summed integral cross section
 for a rotational transition from an initial state $j$ to a final state $j'$
 in the CC formalism is given by
 \begin{equation} \label{cross1}
 \sigma_{j \rightarrow j'}(E_{j}) =
 \frac{\pi }{(2j+1)k^2_{j}} \sum\limits_{J=0}^{J_{\rm max}}(2J+1) 
 \sum\limits_{l=|J-j|}^{J+j} \sum\limits_{l'=|J-j'|}^{J+j'}
 |\delta_{jj'} \delta_{ll'} - S_{jj'll'}^J (E_{j}) |^2 {\rm ,}
 \end{equation}
 where {\bf\textit j} is the rotational angular momentum of the CO molecule,
 {\bf\textit l} is the orbital angular momentum of the collision complex,
 and {\bf\textit J} = {\bf\textit j} $+$ {\bf\textit l} is the total angular momentum.
 $S_{jj'll'}^J$ is an element of the scattering matrix,
 $E_{j}$ is the relative kinetic energy of the initial channel, and
 $k_{j}=\sqrt{2\mu (E - \epsilon_j)}/\hbar$ is the wave-vector of the 
 initial channel, where $\mu$ is the reduced mass of the H-CO system (0.97280~u),
 $E$ is the total energy, $\epsilon_j$ is the rotational energy of CO,
 and $\hbar$ is the reduced Planck constant.
 The CS approximation, on the other hand, reduces the computation expense 
 by neglecting the Coriolis coupling between different values of
 $\Omega$, the projection of the angular momentum quantum
 number of the diatom along the body-fixed axis, and within this formalism 
 the integral cross section is given by
 \begin{equation} \label{cross2}
 \sigma_{j \rightarrow j'}(E_{j}) =
 \frac{\pi }{(2j+1)k^2_{j}} \sum\limits_{J=0}^{J_{\rm max}}(2J+1)
 \sum\limits_{\Omega=0}^{\Omega_{\rm max}}
 (2-\delta_{\Omega0})
 |\delta_{jj'} - S_{jj'}^{J\Omega} (E_{j}) |^2 {\rm ,}
 \end{equation}
 where $\Omega_{\rm max}$ is equal to 
 $0, 1, 2, \ldots$, max($J$,$j$).

State-to-state cross sections for collision energies from 
 10$^{-5}$ to 15,000~cm$^{-1}$ were computed for 
 CO($v=0$,$j$) deexcitation from initial state 
 $j=1-5$, 10, 15, 20, 25, 30, 35, 40, and 45 to all lower $j'$ levels.
 Deexcitation rate coefficients ranging from 1 to 3000~K were obtained by
 averaging the cross sections over a Boltzmann distribution of collision energies,
 \begin{equation} \label{rate1}
 k_{j\rightarrow j'}(T) = \left (\frac{8k_{\rm b}T}{\pi \mu} \right )^{1/2}
 \frac{1}{(k_{\rm b}T)^{2}}\int_{0}^{\infty}\sigma_{j\rightarrow j'}(E_{j})
 \exp(-E_{j}/k_{\rm b}T)E_{j}dE_{j},
 \end{equation}
 where $\sigma_{j\rightarrow j'}$ is the 
 state-to-state rotationally inelastic cross section,
 $E_{j} = E - \epsilon_j$ is the center of mass kinetic energy, and
 $k_{\rm b}$ is the Boltzmann constant.

\section{Results and Discussion}

\begin{figure}
\plotone{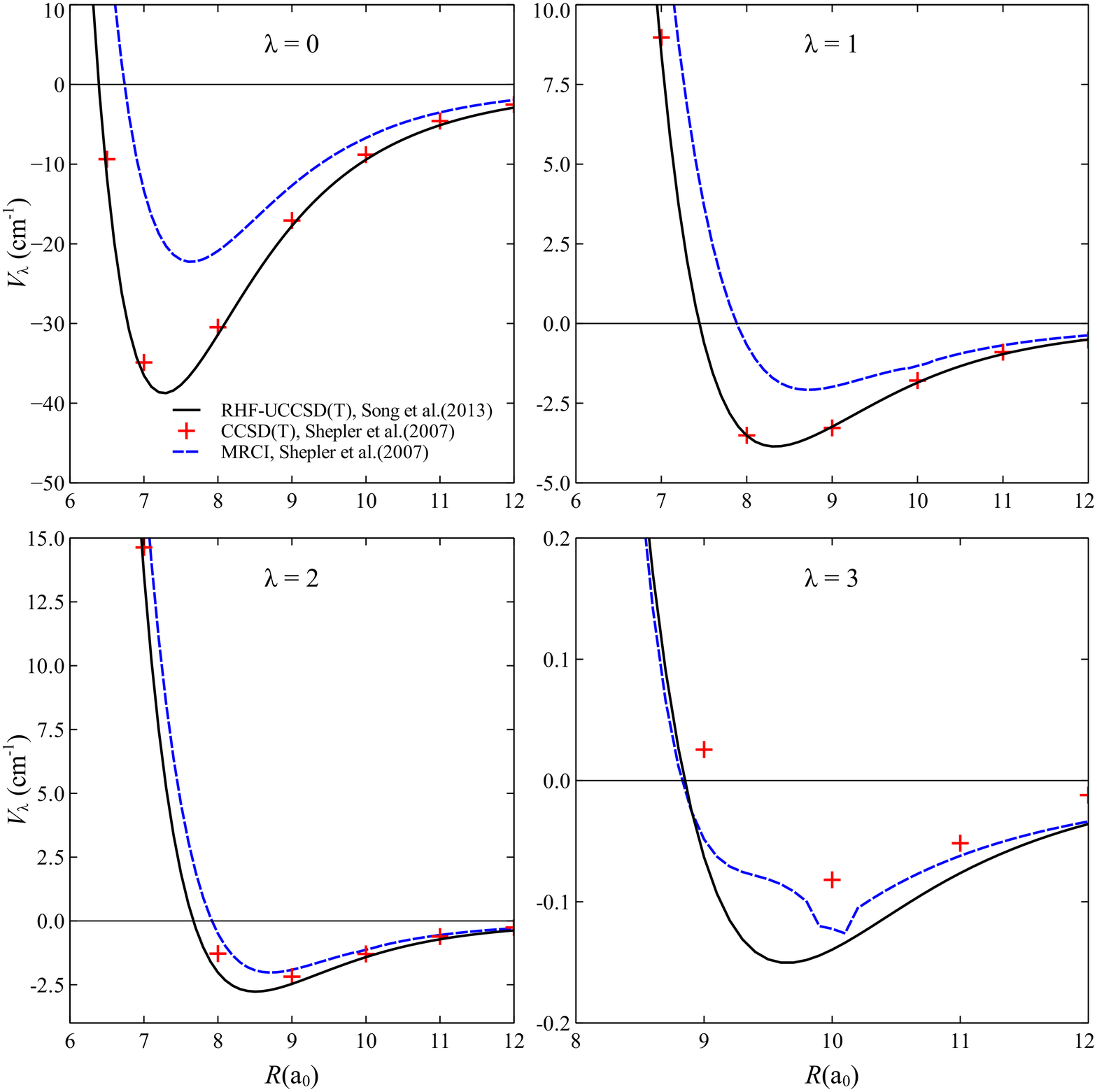}
\caption{The first four Legendre expansion terms 
$v_{\rm \lambda}$($R$) near the van der Waals well of the H-CO PES with 
the CO intermolecular distance fixed at $r = 2.20$~a$_{\rm 0}$ for MRCI 
and the equilibrium bond length $r_{\rm e} = 2.1322$~a$_{\rm 0}$ for the 
CCSD(T) and RHF-UCCSD(T) surfaces.
\label{fig:figr1}}
\end{figure}

\begin{figure}
\plotone{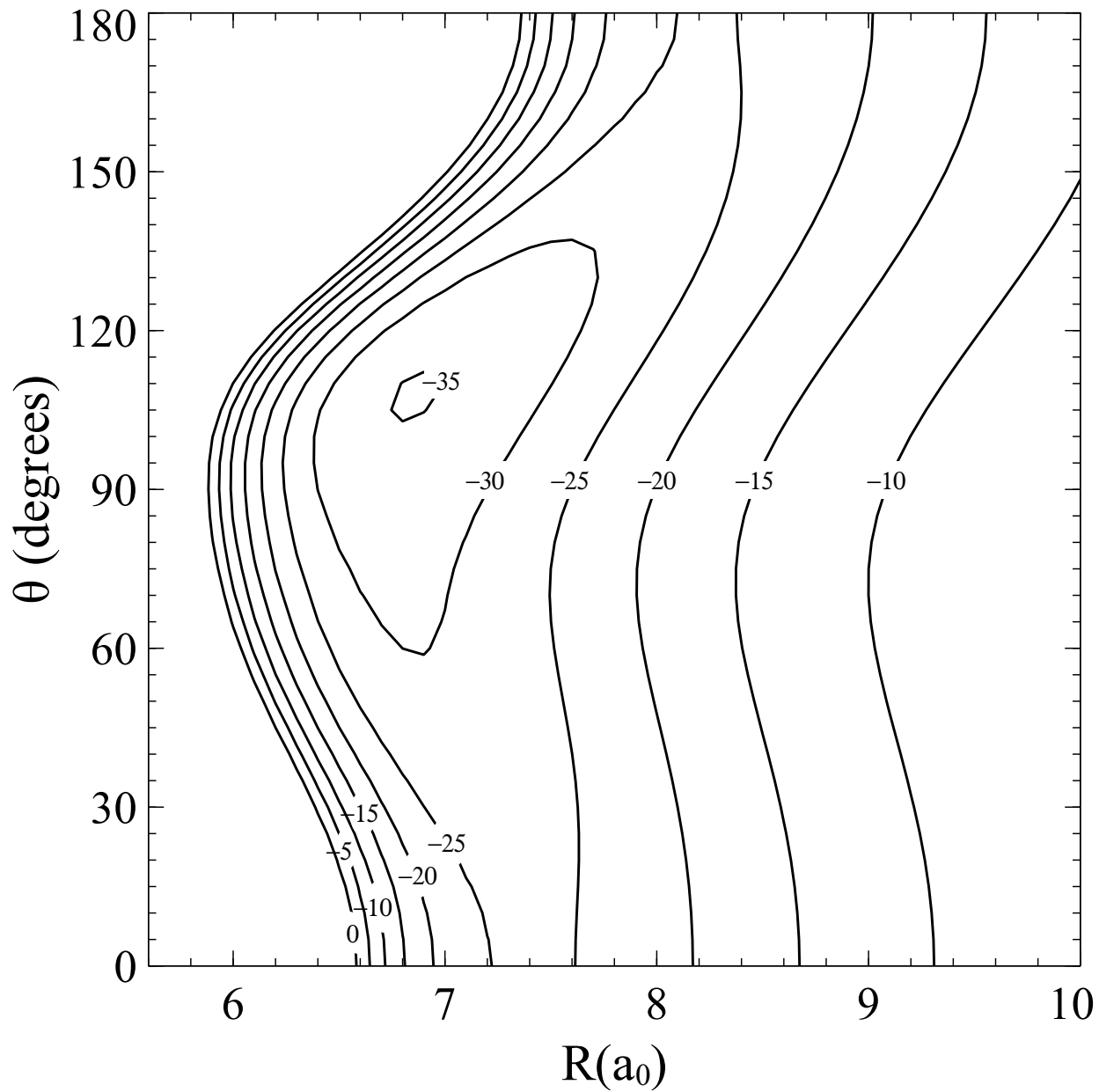}
\caption{
 The long-range behavior of the interaction potential of \citet{song}
 with V$_{\rm HCO}$ in cm$^{-1}$ and $r$ fixed at the equilibrium 
 bond length $r_{\rm e}=2.1322$~a$_{\rm 0}$.
\label{fig:figr2}}
\end{figure}

Before scattering calculations were carried out, we compared the 
 RHF-UCCSD(T) three-dimensional \textit{ab initio} PES of \citet{song} 
 with the two-dimensional rigid rotor MRCI and CCSD(T) 
 surfaces of \citet{shep07}. 
 The lower order Legendre terms 
 of the RHF-UCCSD(T) PES agree well with the CCSD(T) terms;
 the similar high level of theory in both CCSD(T) calculations
 is expected to yield a more accurate potential than the lower-level 
 MRCI calculations (see Figure~\ref{fig:figr1}).
 The higher-order $\lambda$-terms are typically small and decrease
 with $\lambda$. For example, for the RHF-UCCSD calculations shown
 in Figure 1, $v_3 \sim v_0/100$. Further, as the cross sections
 for a direct transition are driven by $v_\lambda$ where
 $|\Delta j| = \lambda$, they are expected to decrease with
 $|\Delta j|$ and as a consequence the uncertainty in the cross
 sections are expected to increase with $|\Delta j|$, though
 multi-step transitions due to smaller $v_\lambda$ terms will also contribute.
 The values of the Legendre expansion terms of the MRCI potential are all
 greater than those of the RHF-UCCSD(T) and CCSD(T) PESs.
 The feature seen in the MRCI results near $R = 10$~a$_0$ is due to the
 poor quality of fitting where the 
 long--range and short--range regions are joined.
 Since the depth and location of the van der Waals well may
 strongly influence rotationally inelastic scattering, 
 the behavior of the three interaction 
 potentials was closely examined in the vicinity of the van der Waals well.
 Figure~\ref{fig:figr2} shows a contour plot of the long-range 
 behavior of the interaction potential of \citet{song}
 with the CO bond length fixed at the equilibrium 
 distance of $r_{e}=2.1322$~a$_{0}$.
 The positions and values of the local and van der Waals 
 minima on the three interaction PESs were calculated and are 
 shown in Table~\ref{table1}.
 The van der Waals well of the MRCI potential is significantly different 
 from that of the CCSD(T) and RHF-UCCSD(T) being both shallower and located at a
 larger internuclear distance.
 The differences in van der Waals well depth and anisotropies of 
 the surfaces lead to differences in the scattering results.

\begin{figure}
\plotone{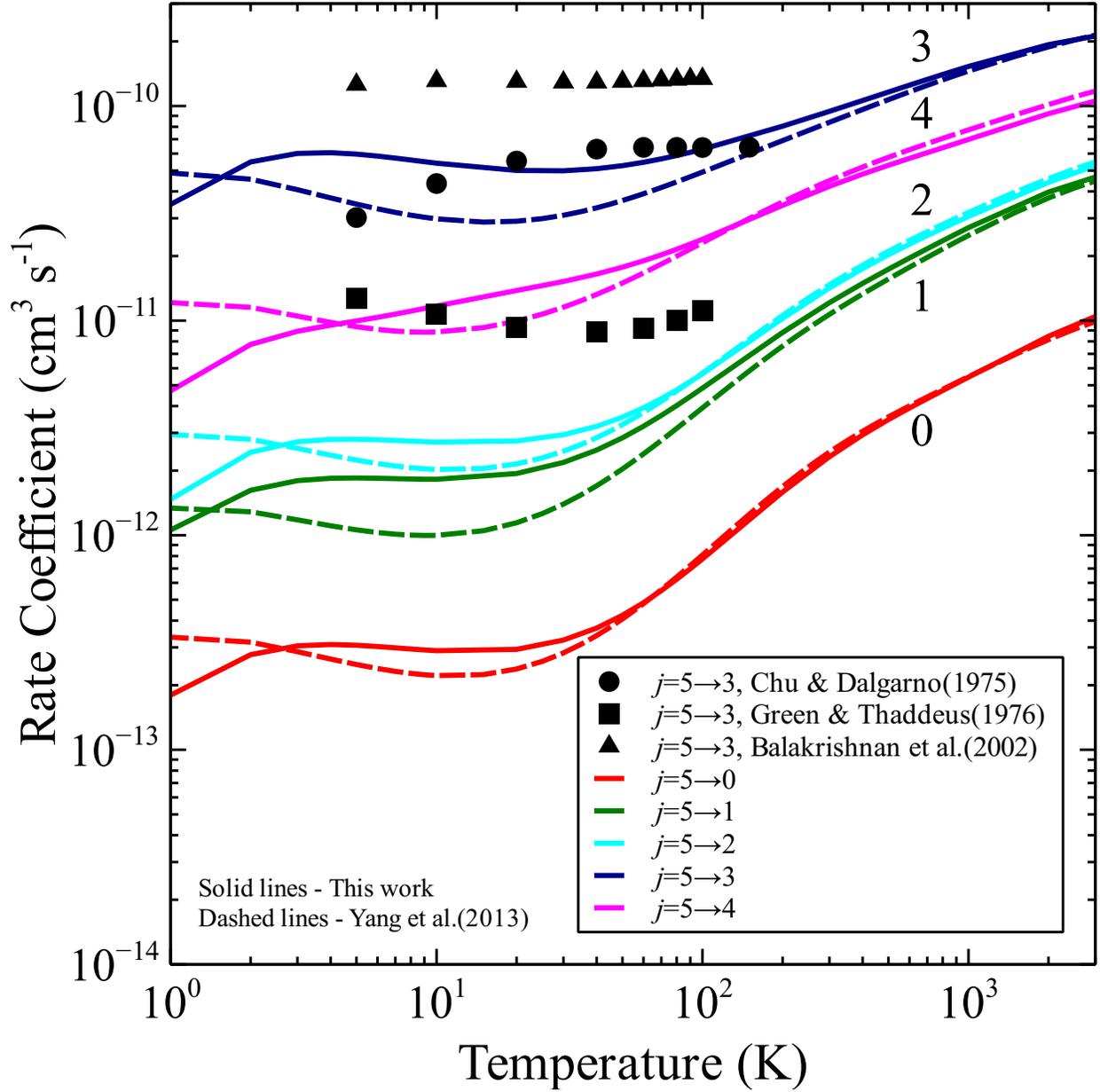}
\caption{State-to-state pure rotational rate coefficients due 
to H collisions from initial state CO($j=5$) to indicated lower 
states $j'$ from this work using the \citet{song} PES, and from 
\citet{yang13} on the MRCI PES of 
\citet{shep07}. Previous rate coefficient 
calculations of \citet{chu75}, \citet{green76}, and \citet{bala02} 
on other surfaces are also shown for the dominant 
transition $j=5\rightarrow 3$.
\label{fig:figr3}}
\end{figure}

Figure~\ref{fig:figr3} shows state-to-state pure rotational H-CO
 rate coefficients from initial state CO($j=5$) to indicated lower 
 states $j'$.
 The largest rate coefficients of CO($5\rightarrow3$) as 
 calculated by 
 \citet{chu75}, \citet{green76}, and \citet{bala02} are shown.
 Rate coefficients from \citet{yang13} on the MRCI PES of 
 \citet{shep07} are also shown.
 The rate coefficients in this work agree well with those of
 \citet{chu75} and \citet{yang13}.
 On average, rate coefficients of \citet{green76} are less than 
 those of this work by around an order of magnitude,
 while the values from \citet{bala02} are around twice as much as ours.
 Although the CO($5\rightarrow3$) transition is highlighted here,
 the results are typical for other state-to-state transitions
 as well.

\begin{figure}
\plotone{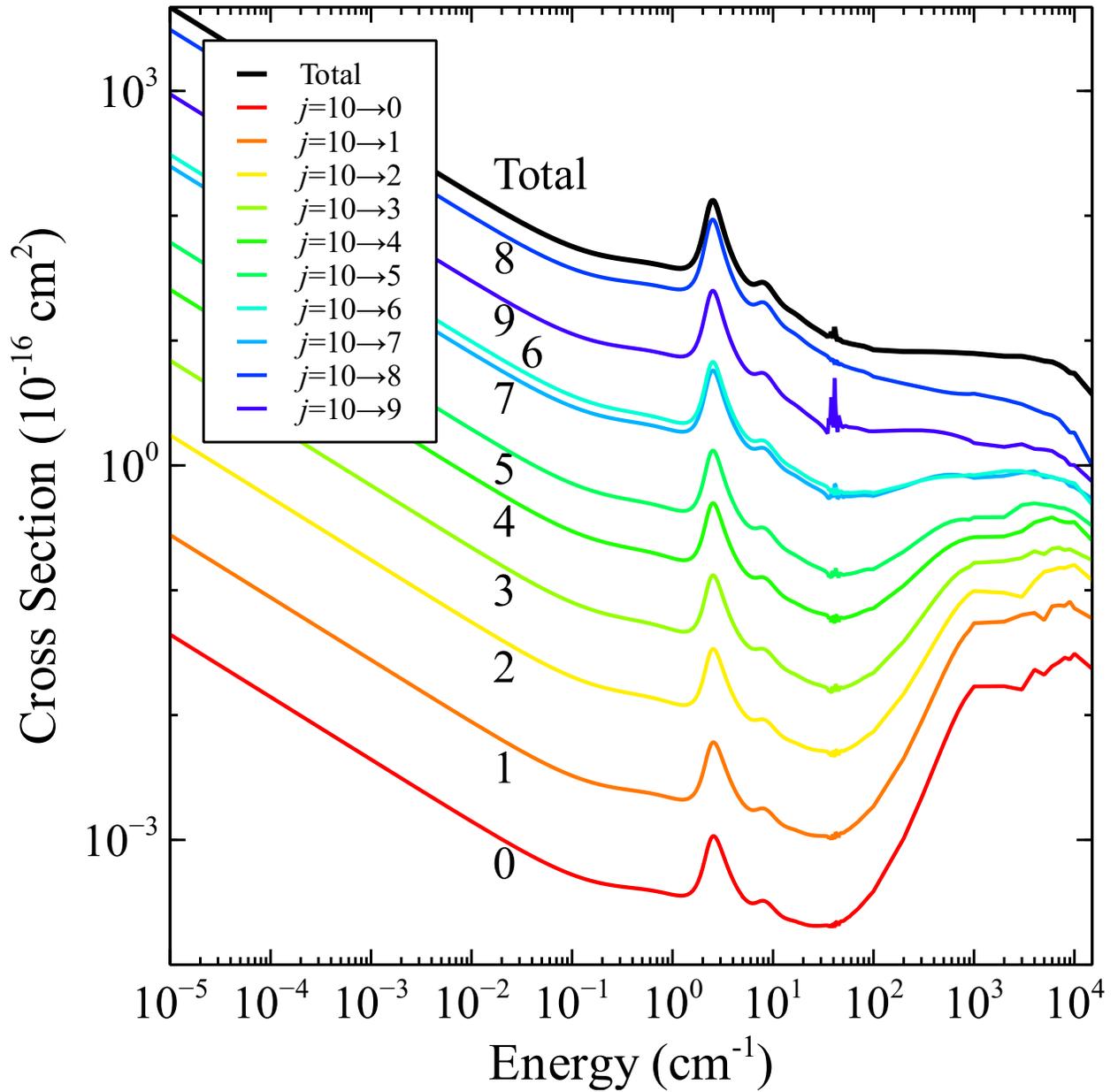}
\caption{State-to-state pure rotational deexcitation cross sections 
due to H collisions from initial state CO($j=10$) to all lower states $j'$.
\label{fig:figr4}}
\end{figure}

\begin{figure}
\plotone{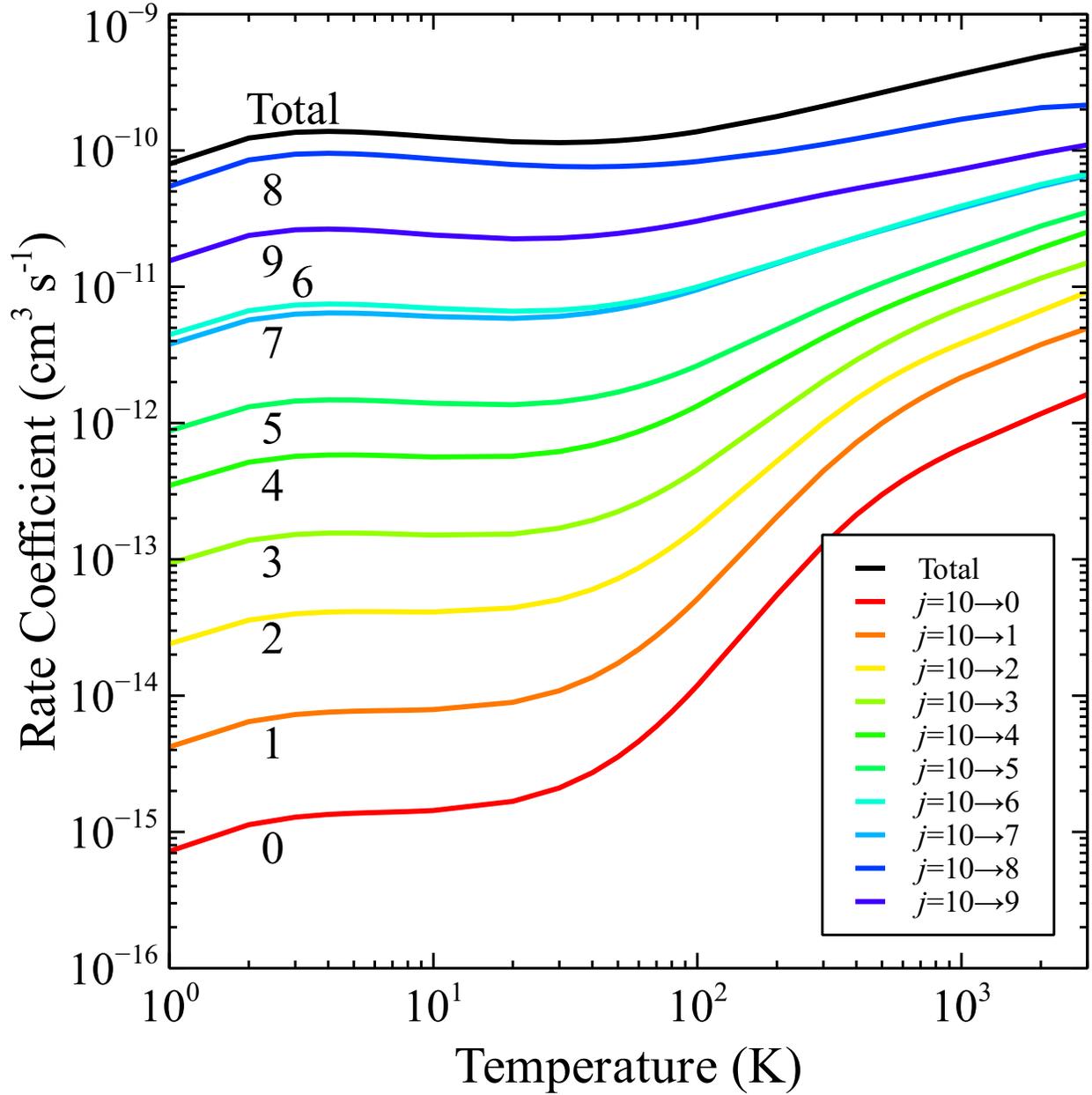}
\caption{State-to-state pure rotational deexcitation  
rate coefficients due to H collisions from initial state CO($j=10$) to
all lower states $j'$.
\label{fig:figr5}}
\end{figure}

Figures~\ref{fig:figr4} and \ref{fig:figr5} present
 sample results from our computations of cross sections and rate 
 coefficients, respectively, from initial state $j$=10 to all final states $j'$.
 In general, the cross sections for $j'$=0 are smallest and 
 then increase with increasing $j'$ up to the dominating
 transition where $|\Delta j|=|j'-j|=2$ after which they decrease.
 Since CO is near-homonuclear, odd-$\Delta j$ transitions are 
 suppressed and the transitions follow an even-$\Delta j$ propensity.
 At 1000~cm$^{-1}$, the difference between the CC and CS cross sections 
 is less than $\sim$5\%, except for the largest changes in $\Delta j$ 
 where the differences are typically $\sim$10\%.

\begin{figure}
\plotone{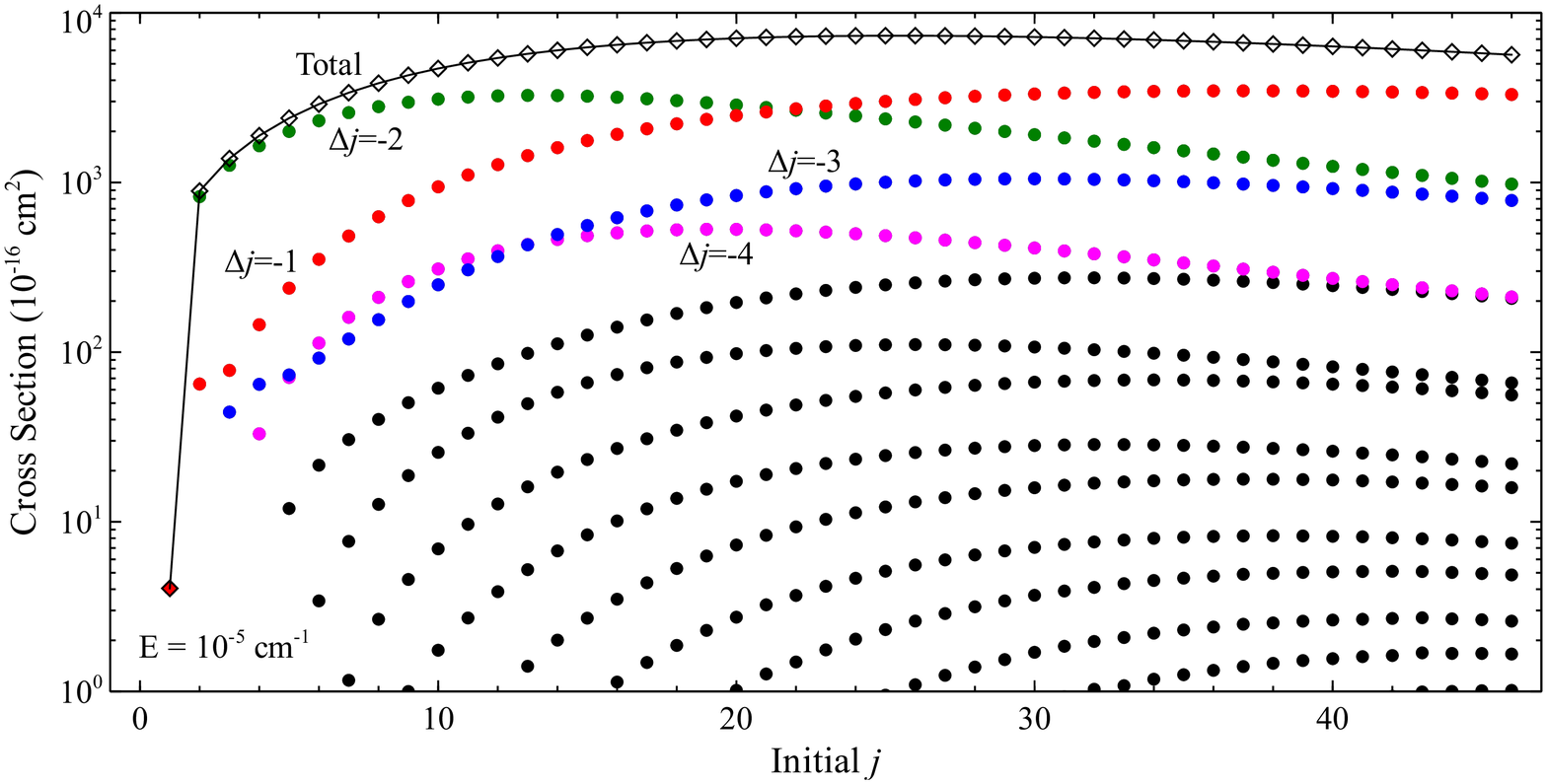}
\caption{State-resolved cross sections at 10$^{-5}$ cm$^{-1}$ as a function of 
increasing $j$ for the H-CO system. Circles indicate state-to-state 
cross sections while diamonds indicate total quenching cross sections.
Each series (circles) correspond to different values of $\Delta j$, decreasing
from the top at the far right as $\Delta j= -1, -2, -3, -4, ...$ (red, green,
blue, magenta).}
\label{fig:figr6}
\end{figure}

\begin{figure}
\plotone{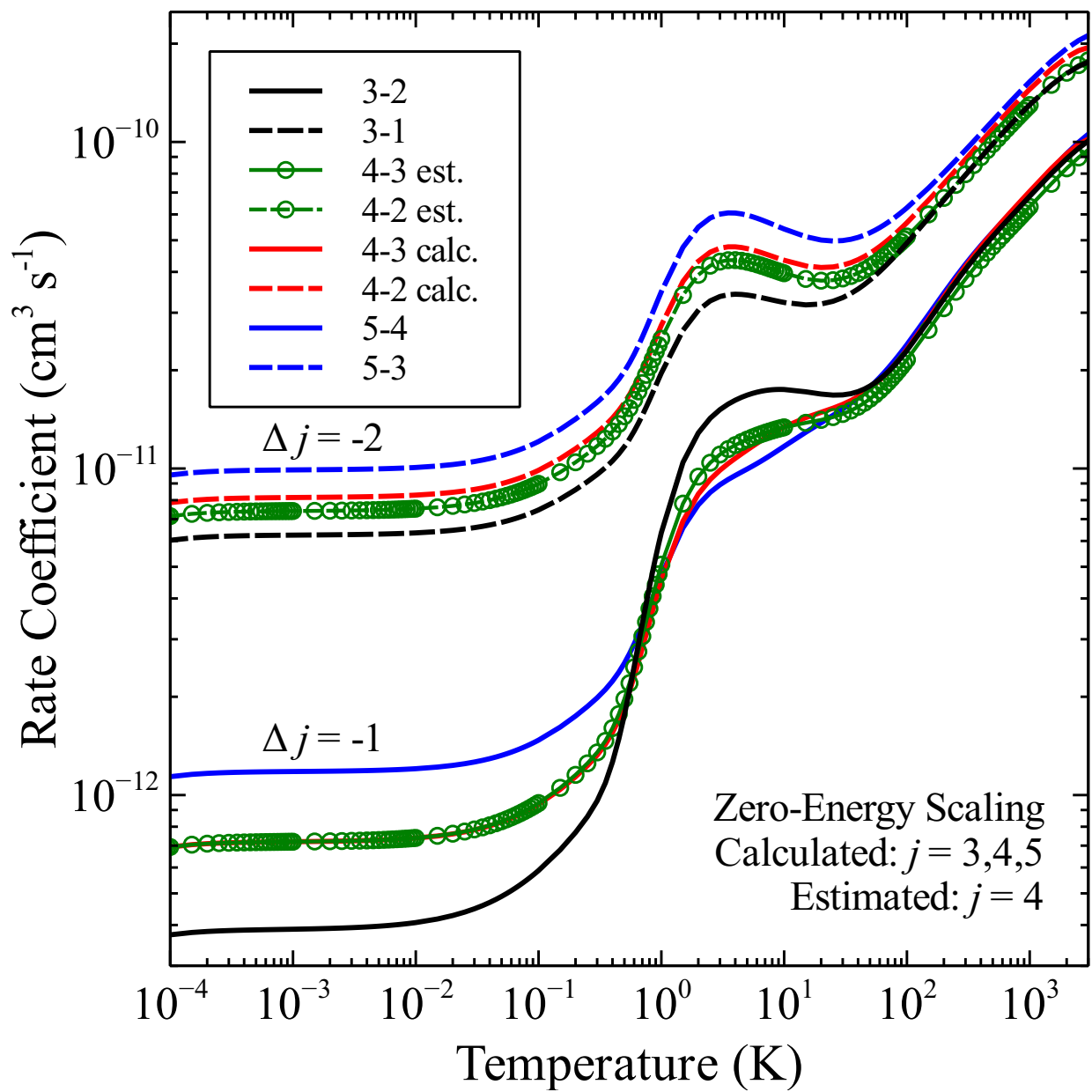}
\caption{A comparison of the zero-energy scaling technique with explicit
calculations of the largest deexcitation 
rate coefficients for initial state $j=4$. Solid lines indicate $\Delta j=-1$
while dashed lines are $\Delta j=-2$. Initial states are: $j=3$~(black), 
$j=4$ estimated from Equation~\ref{zest}~(green with symbols), $j=4$ explicitly 
calculated~(red), and $j=5$~(blue).
\label{fig:figr7}}
\end{figure}

While the quenching from selected high rotational states is explicitly 
 calculated, a zero-energy scaling technique~(ZEST) is used to predict 
 state-to-state rate coefficients for all intermediate states.
 Figure~\ref{fig:figr6} shows the total quenching and state-to-state cross sections 
 calculated at 10$^{-5}$~cm$^{-1}$ as a function of increasing $j$.
 The total quenching rate coefficient in the ultracold limit
 (collision energy of 10$^{-5}$~cm$^{-1}$) is fairly constant
 for $j>10$. Note that $\Delta j = -2$ dominates for low initial $j$, 
 as expected, but at $j\sim22$, $\Delta j = -1$ begins to dominate. 
 From these ultracold state-resolved quenching rate coefficients, 
 we estimated the rate coefficients for all temperatures following the
 procedure introduced in \citet{yang06}, but modified here. 
 The unknown rate coefficient for a transition
 from $j$ to $j-\Delta j$ for any temperature $T$, given by $k(j,j-\Delta j; T)$,
 can be estimated based on the rate coefficients from calculated transitions 
 for other initial rotational states, for example, larger $j$ 
 (i.e., above, $j_A$) or smaller $j$ (i.e., below, 
 $j_B$)  [$k(j_A,j_A-\Delta j; T)$ and 
 $k(j_B,j_B-\Delta j; T)$, respectively]
 and calculated ultracold ($T\sim0$)
 rate coefficients for the transitions above, below, and the desired
 according to 
 \begin{equation} \label{zest}
 k(j,j- \Delta j;T) = k(j,j- \Delta j;T=0) 
 \frac{w_A k(j_A,j_A- \Delta j;T) + w_B k(j_B,j_B- \Delta j;T)}
 {w_A k(j_A,j_A- \Delta j;T=0) + w_B k(j_B,j_B- \Delta j;T=0)}.
 \end{equation}
 Weights $w_A$ and $w_B$ for the calculated rate coefficients 
 were determined by considering the change in initial $j$ in the 
 desired transition when compared to the states above and below 
 according to the equations
 \begin{equation} \label{weight}
 w_A = \frac{j-j_B}{j_A-j_B}
 \qquad
  w_B=\frac{j_A-j}{j_A-j_B}.
 \end{equation}
 For example, the predicted $j=16$ rate coefficients 
 used calculated transitions from
 $j=20$ and $j=15$ and weights $w_A=0.2$ and $w_B=0.8$.
 The cross sections at ultracold energies 
 are easily calculated and simply multiplied by the velocity of the 
 system (obtained from the kinetic energy of the collision) to 
 obtain the $T\sim0$ rate coefficients. A comparison of this zero-energy
 scaling technique with explicit calculations is given in Fig.~\ref{fig:figr7}
 for initial state $j=4$. The predicted rate coefficients agree well 
 with the calculated values, although in this case transitions from 
 $j=3$ and $j=5$ are equally weighted and a slight underestimation 
 of the $j=4-2$ transition occurs.
 Given the high accuracy of the current PES, any uncertainty remaining
 in the computed cross sections is likely due to the
 adoption of the CS approximation,
 the truncation in the Legendre
 expansion of the PES to $\lambda_{\rm max}=20$ and the uncertainty in
 $v_\lambda$ for large values of $\lambda$, and limited basis set sizes for
 collision energies above 1000 cm$^{-1}$. Otherwise, the interpolation
 formula introduces the largest source of uncertainty.

\section{Astrophysical Applications}

The cross sections and/or rate coefficients 
 calculated in this work are available online.\footnote{Rate coefficient data in
 the Leiden Atomic and Molecular Database \citep[LAMDA,][]{sch05} format,
 as well as cross section data can be obtained at 
 www.physast.uga.edu/amdbs/excitation/.}
 These high-$j$ pure rotational rate coefficients are especially useful 
 as input data for codes developed to solve for level populations.
 One such code is RADEX \citep{tak07}, which can perform a non-LTE 
 analysis of interstellar line spectra.
 The reliable rate coefficients calculated in this work also will help 
 extract more accurate astrophysical conclusions from current
 models.
 Recently, \citet{thi13} modeled protoplanetary disks (PPDs) adopting the 
 rate coefficients computed by \citet{bala02}.
 These rate coefficients are based on an inaccurate interaction surface,
 so the conclusions of \citet{thi13} may be compromised.
 Furthermore, \citet{thi13} extrapolated the rate coefficients to higher 
 temperatures.
 Our rate coefficients extend the range in temperature from previous studies
 up to 3000~K; therefore extrapolation in this case would not be necessary.
 It is expected that our reliable and comprehensive rate 
 coefficients would lead to more accurate astrophysical models of PDRs, PPDs, and
 other molecular environments. 

\begin{figure}
\epsscale{0.9}
\plotone{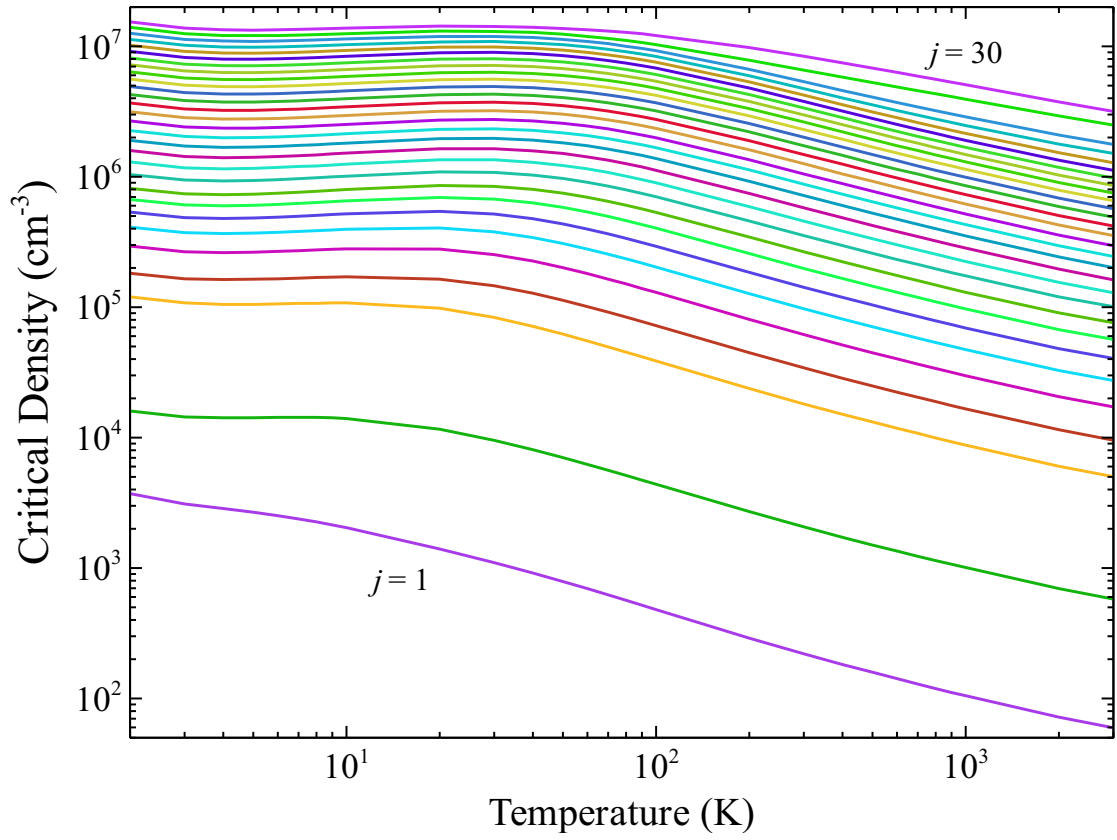}
\caption{Critical densities for CO($v=0,j$) due to H
collisions as a function of gas temperature $T$.}
\label{fig:figr8}
\end{figure}

In the PPD models of \citet{thi13}, they note that pure rotational transitions of CO probe
the entire disk, while the warm inner regions are probed by rovibrational transitions.
Further, the abundance of atomic hydrogen is high in the CO line-emitting region. In
particular, it is typically greater than 10$^6$ cm$^{-3}$ throughout the disk, except
near the mid-plane and at the disk surface. Assuming that H collisions and
spontaneous emission, with transition probability $A_{j\rightarrow j^\prime}$,
 dominate the CO rotational populations, the critical
density for each rotational level $j$ can be estimated following 
\citet{ost06} with the relation
\begin{equation}
n^{\rm cr}_j = \frac{\sum\limits_{j' < j} A_{j\rightarrow j'}} 
{\sum\limits_{j'\neq j} k_{j \rightarrow j'}} {\rm ,}
\end{equation}
and are displayed in Figure~\ref{fig:figr8} \citep[a similar figure
for para-H$_2$ collisions can be found in][]{yang10}. Except near the inner
disk where temperatures exceed 500 K and the density is greater than 
10$^8$ cm$^{-3}$, levels for $j\gtrsim 15$ will not be in LTE. Infrared and
UV-fluorescence pumping may also contribute resulting in supra-thermal
CO rotational populations.

\section{Conclusion}

The three-dimensional H-CO potential energy surface of \citet{song} 
 was used to perform quantum scattering calculations for rotational 
 deexcitation transitions of CO induced by H. 
 State-to-state cross sections for collision energies from 
 10$^{-5}$ to 15,000~cm$^{-1}$ and rate coefficients for temperatures 
 ranging from 1 to 3000~K were computed for CO pure rotational deexcitation.
 Not surprisingly, previous scattering results using a PES based on 
 the CCSD(T) level of theory produced similar rate coefficients, but considered
 only low rotational excitation states.
 The MRCI surface of \citet{shep07} with a less sophisticated level of theory
 also produced similar results as presented in \citet{yang13},
 while the scattering results of \citet{bala02} are deemed to be unreliable
 due to an inaccurate surface.
 The H-CO pure rotational rate coefficients presented here can be used to 
 aid astrophysical modeling and
 they both extend the temperature range and angular momentum quantum number
 of reported H-CO rate coefficients.
 Further, while  calculations of deexcitation 
 from excited rovibrational levels have
 recently been performed \citep{song15a,song15b},
 experimental data on low-temperture rotational and
 vibrational inelastic rate coefficients would be highly desirable.

\acknowledgments

We would like to thank Shan-Ho Tsai of The Georgia Advanced Computing 
 Resource Center at the University of Georgia for computational assistance and
 the Computer \& Communications Department at Radboud University for
 computing resources.
 This work was partially supported by NASA grant NNX12AF42G.


\clearpage

\begin{table}
\begin{center}
\caption{Minima of the interaction PESs.
\label{table1}}
\begin{tabular}{lllll}
\tableline \tableline
PES & $r_{\rm CO}$(a$_{\rm 0}$) & $V_{\rm min}$(cm$^{-1}$) & $R$(a$_{\rm 0}$) & 
$\theta$ (degrees)\\
\tableline
Global minimum:\tablenotemark{a} &&&& \\
MRCI\tablenotemark{b} & 2.20 & -6934.052 & 3.01 & 145.5\\
CCSD(T)\tablenotemark{b} & 2.1322 & -5817.161 & 3.00 & 144.7\\
RHF-UCCSD(T)\tablenotemark{c} & 2.1322 & -5832.994 & 3.01 & 144.8\\
RHF-UCCSD(T)\tablenotemark{c} & 2.20 & -7256.805 & 3.02 & 144.8\\
\tableline
van der Waals minimum: &&&&\\
MRCI & 2.20 & -19.934 & 7.21 & 110.8\\
CCSD(T) & 2.1322 & -34.677 & 6.87 & 108.1\\
RHF-UCCSD(T) & 2.1322 & -35.269 & 6.86 & 108.0\\
RHF-UCCSD(T) & 2.20 & -36.296 & 6.84 & 108.5\\
\tableline \tableline
\end{tabular}
\tablenotetext{a}{The term ``global" is used loosely here since $r_{\rm CO}$ 
 has been frozen at an unoptimized distance to ease direct comparison
 with the previous 2D surfaces.}
\tablenotetext{b}{\citet{shep07}}
\tablenotetext{c}{\citet{song}}
\end{center}
\end{table}

\clearpage

\clearpage

\clearpage

\clearpage

\clearpage

\clearpage

\clearpage

\clearpage

\end{document}